\documentclass{article}
\usepackage{ijcai24}

\usepackage{times}
\usepackage{soul}
\usepackage{url}
\usepackage[hidelinks]{hyperref}
\usepackage[utf8]{inputenc}
\usepackage[small]{caption}
\usepackage{graphicx}
\usepackage{float}
\usepackage{amsmath}
\usepackage{amsthm}
\usepackage{booktabs}
\usepackage{algorithm}
\usepackage{algorithmic}
\usepackage{xcolor}
\usepackage[switch]{lineno}

\title{Exploring Prompt Engineering Practices in the Enterprise}

\author{
Michael Desmond\thanks{Equal contribution} 
\and 
Michelle Brachman\footnotemark[1]
\affiliations
IBM Research\\
\emails
mdesmond@us.ibm.com,
michelle.brachman@ibm.com
}

\begin{document}

\maketitle

\begin{abstract} 
Interaction with Large Language Models (LLMs) is primarily carried out via prompting. A prompt is a natural language instruction designed to elicit certain behaviour or output from a model. In theory, natural language prompts enable non-experts to interact with and leverage LLMs. However, for complex tasks and tasks with specific requirements, prompt design is not trivial. Creating effective prompts requires skill and knowledge, as well as significant iteration in order to determine model behavior, and guide the model to accomplish a particular goal. We hypothesize that the way in which users iterate on their prompts can provide insight into how they think prompting and models work, as well as the kinds of support needed for more efficient prompt engineering. To better understand prompt engineering practices, we analyzed sessions of prompt editing behavior, categorizing the parts of prompts users iterated on and the types of changes they made. We discuss design implications and future directions based on these prompt engineering practices.
\end{abstract}

\section{Introduction}

With the emergence of instruction tuning and alignment techniques, prompting is the primary mode of interaction with LLMs ~\cite{liu2023pre}. 
In the enterprise, developers and AI practitioners try to develop prompts to automate knowledge tasks of varying complexity, in order to improve the efficiency of the organization (and harness the value of AI) ~\cite{cambonearly}. 
This includes tasks like summarizing a document or transcript, generating code or other structured output, and content-grounded Q\&A~\cite{ritala2023transforming}. 
Regardless of task, the prompts typically include various components, sometimes include embedded examples, and require outputs that fit particular requirements and high levels of accuracy~\cite{braun2024can}. 
Further, enterprise contexts may call for the use of models that are more challenging to prompt for a variety of reasons, like cost and specialization.
It's not always clear if an LLM is capable of performing a given task, and significant effort is involved in developing and refining prompts in an attempt to find out. Some studies have begun to explore prompt engineering behaviors~\cite{zamfirescu2023johnny} and prompt structures~\cite{braun2024can}. However, prompt engineering is still a new discipline and behaviors likely differ across contexts and domains. We currently know very little about how practitioners in the enterprise edit prompts over time~\cite{schmidt2023cataloging}. 

We were interested in observing how practitioners edit and refine prompts as a means of exploring and controlling LLM behavior.
We believe this process can help us  understand how practitioners understand and interact with LLMs, and what kinds of tools would help to make the discovery and engineering process more efficient. 
To address these questions, we collected a large amount of interaction data from an enterprise-scale LLM prompting environment. 
This data captured the prompts that users applied to a set of hosted LLMs, and allowed us to study the editing and refinement process that took place over time. 
We analyzed a sample of 57 users' prompting sessions for qualitative analysis across an array of use cases. We captured the prompt component that the user edited, the type of edit applied, as well as whether an edit was un-done or re-done. 

Our contributions include a large-scale analysis of prompt editing practices across varying use cases in an enterprise context and corresponding design implications.
Overall, we found that prompt editing sessions were primarily made up of prompt edits mixed with model switches. The most commonly edited prompt component was the context, followed by task instructions and labels and the most common edit type was modification, where the meaning stays the same. 


\section{Related Work}


\subsection{Prompting Practices}
Prompt engineering and design is the process of formulating the natural language text that is input to a Large Language Model. Researchers study prompt engineering both in terms of technical strategies that improve performance, as well as human-centered study of how users actually interact with prompts. On the technical side, researchers have developed a variety of strategies to improve prompt performance, such as few-shot learning~\cite{brown2020language}, chain of thought~\cite{wei2022chain}, and automated prompt generation strategies~\cite{wang2023promptagent,melamed2023propane}. While these strategies may improve performance, they may be context specific and do not come with guarantees. Our work contributes to the relatively new area of understanding prompting practices from a human-centered perspective. 

Several studies have identified challenges for non-experts in prompt engineering, finding that they use trial and error~\cite{dang2022prompt}, are willing to reformulate their prompts using support~\cite{bodonhelyi2024user}, generalize too much from individual instances, and expect LLMs to act like humans~\cite{zamfirescu2023johnny}. 
A review of literature categorized types of dissatisfaction users have when interacting with LLMs, such as issues with the response format and attitude or the intent understanding. They also investigated users' tactics when repairing incorrect interactions with ChatGPT, such as making an intent more concrete, pointing out errors, or adapting the task ~\cite{kim2023understanding}. A small body of work has begun to define ``prompt patterns'' for ChatGPT, which include both components of prompts as well as prompt modification types~\cite{white2023prompt,schmidt2023cataloging}. Our work builds on this work by analyzing a large set of prompts across use cases and models, focusing on prompt editing practices and capturing their usage across an enterprise dataset. 

As prompting is a still emerging practice, many resources have been created to support users in designing and iterating on their prompts, from blogs to courses to research papers. Often, research papers for particular models will provide insight into prompting strategies that work well. Researchers have further focused specifically on prompt practices, identifying a taxonomy of prompt design dimensions~\cite{braun2024can}, design considerations for prompting and a typology of prompting methods~\cite{liu2023pre}. More closely aligned with online resources for prompting, some work provides general instructions, like that prompts should include examples, be specific, and ask for multiple options~\cite{lin2023ten}.

Our work contributes to the understanding of prompt engineering practices through an analysis of LLM prompt iteration in an enterprise context. We analyze both the part of the prompt users modify as well as the type of edit they make to the prompt across a dataset of prompt sessions. This provides significant insight into real prompting patterns, as well as their prevalence.





\subsection{Mental Models and Repair} 
We expect that the types of prompting practices used will provide insight into users' mental models, or understanding of, how to best prompt Large Language Models. Mental models, and methods for supporting mental models have been studied extensively for traditional AI and Machine Learning methods~\cite{bansal2019beyond,gero2020mental}.
The recent advancement of LLMs raises new questions about how to support users' mental models, as there are new challenges in existing methods, like transparency, for LLMs~\cite{liao2023ai,bommasani2023foundation}. Yet, we know little about users' current assumptions about the best way to prompt LLMs.

Several studies have investigated users' mental models and usage of LLMs and generative AI for code. Users found the open-ended nature of natural language code generation to be challenging~\cite{liu2023wants}, wishing for constraints~\cite{jiang2022discovering} or information about what kinds of inputs generative AI systems can handle~\cite{sun2022investigating}. One study revealed a set of repair strategies used during generative program synthesis, including rewording, expanding scope, and changing model parameters~\cite{jiang2022discovering}. Another study explored repair strategies for code generation, such as elaboration, language restructuring, and intent shaping~\cite{liu2023wants}. While related, these studies focused only on code generation tasks and only on a user query. Our work contributes an analysis of how people edit prompts across a variety of use cases in an enterprise context.

\section{Methods}

\subsection{Data Collection}
We collected our dataset of prompts from an internal enterprise platform for experimentation and development with LLMs. The platform has a web-based user interface (UI) for prompt engineering wherein users can input their prompt text, submit the prompt, and the LLM output is generated and appended to their prompt. Users can also modify a variety of generation parameters such as decoding strategy, temperature, repetition penalty etc. The tool is open to internal use across the company and thus could include users of varying prompt engineering experience and expertise. It allows users to use a variety of LLMs, including both open-source models (like llama-2 and flan) and proprietary models. Each datum includes the following information:
\begin{itemize}
    \item User Id
    \item Time stamp
    \item Prompt text
    \item Target language model 
    \item Generation parameters 
\end{itemize}

We collected the above data for 1,712 users from August 31, 2023 to September 20, 2023. The dataset was anonymized before data collection. From the raw data we calculated each users prompt editing history, and then split it into editing ``sessions'' based on breaks in time of at least 20 minutes between subsequent records.
We sampled from the dataset first in order to conduct exploratory analysis, and then again to carry out our final analysis.  

For our exploration dataset, we aimed to sample a diverse set of prompt editing behavior. To do this, we sampled: 1) a random set of 40 users' sessions, selected based on the number of prompts they had run on the platform by quartile (10 who had less than 4 prompts, 10 who had between 4 and 11 prompts, 10 who had between 12 and 29 prompts and 10 who had at least 30 prompts), and 2) a random sample of ten sessions across the top 20 users, as these users had many more submitted prompts than the other users selected (over 300) and may have different practices.
For our inter-rater agreement and final analysis, we sampled a new set of sessions distinct from the exploration dataset, to ensure that our analysis would hold on a new set of data. Based on our exploratory analysis, we discovered that short prompting sessions rarely included depth in prompt iteration, so we aimed to analyze a broad set of sessions of moderate length with a high edit rate. We first randomly sampled one session per user of length at least 20 prompts in which at least 75\% of prompts had a prompt edit (as opposed to a model or model parameter change), leading to a set of 330 sessions. To select a manageable final dataset, we randomly selected 100 of these sessions. From the 100 sessions, we rejected sessions in which prompts were not in English or in which users tried many sample prompts rather than coherently iterating toward a specific task, ending with a final set of 57 sessions. For our qualitative analysis, we clipped any sessions longer than 50 edits to 50 edits.

\subsection{Qualitative Analysis}
We performed a qualitative analysis of prompt edits and practices in three stages: 1) exploratory code development, 2) inter-rater agreement, and 3) coding. We developed and used a web-based tool that highlights the differences between pairs of successive prompts in order to label the edits between them. For each pair of successive prompt variants, our qualitative analysis captures the edits that occurred from the first variant to the second variant.

The aim of the exploratory code development was to establish and label the types of prompt edits. The two authors looked at our exploratory data set independently, taking notes and forming codes. They then came together, merging their codes, creating definitions for the codes and providing examples of the codes. The resulting code book has categories of codes for: the prompt component that was edited (see Table~\ref{tab:code-part-freq}) and the type of edit that occurred (see Table~\ref{tab:code-change-freq}).

\begin{table*}[t]
\begin{tabular}{ p{3cm}  p{11cm} p{2cm} }
 \toprule
Edit Type &  Description  & Count \\ \hline
 modified & Part of the prompt is edited but has the same meaning or similar meaning, such as rephrasing, simplifying, or grammatical changes. & 602 \\
 added  & New text is appended to the prompt. & 479 \\
  changed & Part of the prompt is edited or replaced such that it has a new meaning.  &306 \\
 removed  & Text is removed from the prompt. & 258 \\
 formatted & Part of the prompt is re-positioned or white space is moved. & 80 \\
 other & not described above & 20\\
 \bottomrule
\end{tabular}
\caption{Codes and descriptions for the types of edits made to prompts. }
\label{tab:code-change-freq}
\end{table*}

\begin{table*}[h]
\begin{tabular}{ p{4cm}  p{8cm} p{4cm}}
 \toprule
Prompt Component &  Description &Example\\ \hline
 instruction:task & The goal or aim of the prompt and other details related to the requested output. & \textit{Answer a question based on a document}; \textit{Summarize a document}\\
 instruction:persona & The type of person or role the LLM should take on when generating the output. & SQL Expert; AI Assistant\\
 instruction:method &A description of the process the LLM should go through to generate the output. & Step-by-step\\
 instruction:output-length & A description of the length of the generated output. & \textit{50 words}; \textit{concise}\\
 instruction:output-format & What form the output should take on. & \textit{JSON}; \textit{a paragraph }\\
 instruction:inclusion & Inclusion prompt elements describe what an output should \textit{or should not} include. & Explanation, specific information from a provided document \\
 instruction:handle-unknown & Handle-unknown is a description of what the output should be if the LLM is missing the knowledge needed to generate the requested output. &\textit{If you don't know, respond with [...]} \\
 label & Labels include text in the prompt that identifies an element of a prompt.  &\textit{Instruction:}; \textit{$<$Context$>$  $</$Context$>$}\\  
 context & We define context as including examples, documents for grounded responses, and input queries. &  \\  
 other & not described above &\\
 \bottomrule

\end{tabular}
\caption{Codes, descriptions and examples for prompt components users edited.}
 \label{tab:code-part-freq}
\end{table*}

The authors then performed an inter-rater agreement task, in which they looked at a subset of 6 sessions (11\% of the data), each labeling the same set independently with one or more of the established codes for both prompt component and edit type. Due to complexities of multiple codes per edit and multiple categories of codes, we calculated inter-rater reliability by calculating the average of percentage of matching codes over all edits, reaching 70\%. We accept this level of inter-rater agreement due to the complexity of labeling prompt components and prompt edit types, as well as the number of codes in our code book. Finally, the authors divided and independently coded the remaining data, for a total of 57 sessions, comprised of 1523 individual prompt edits. 

In addition to our codes, we captured edit rollbacks and session use cases. A rolled back edit meant that it was un-doing or re-doing a previously observed edit. Two authors also recorded the use case of each prompt session based on the task instructions and discussed any disagreements or ambiguities.

\section{Results}

The results are broken into two sections. First we present a high level quantitative analysis of the observed prompt editing sessions. Secondly, we report more detailed results based on our review and annotation process, including qualitative observations.

\subsection{Prompt Engineering Sessions: High-level Editing Analysis}

\begin{figure}[tb]
    \centering
    \includegraphics[width=0.4\textwidth]{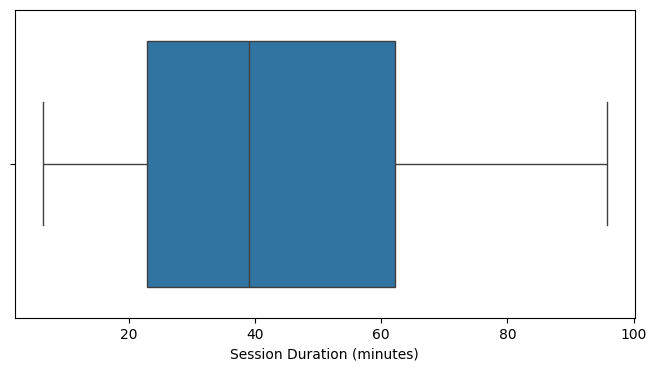}
    \caption{Duration of observed prompt editing sessions in minutes.}
    \label{fig:session_duration}
\end{figure}

Prompt editing sessions were often relatively long. The mean time spent on an editing session was 43.4 minutes (SD=24.5), and the median duration was 39 minutes (See fig. \ref{fig:session_duration} for the distribution). In terms of the time spent editing individual prompts (the time between successive prompt submissions to the platform), we observed that users spent a mean time of 47 seconds (SD=40) editing (not adjusted for inference/text generation time), with a median of 32 seconds. For robustness we winzorized the upper and lower 1\% of prompt editing duration data before reporting statistics. 

\begin{figure}[H]
    \centering
    \includegraphics[width=0.4\textwidth]{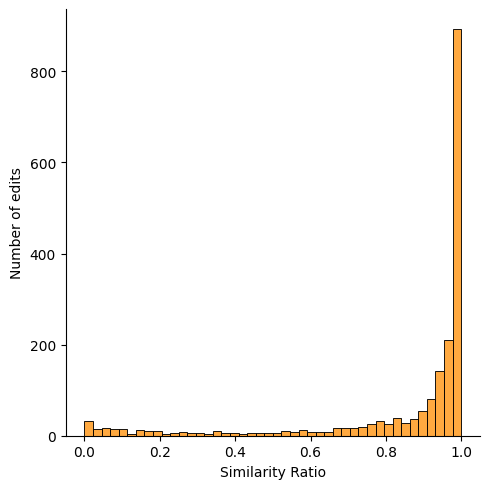}
    \caption{The size of change between successive prompts, represented as a similarity ratio. Values closer to 1 indicate similarity between successive prompts, 1 being an exact match, while a ratio of 0 indicates nothing in common. }
    \label{fig:edit_ratios}
\end{figure}

The mean per session \textit{edit rate} observed was .86 (SD=.13), and the median was 0.9, indicating that close to 9/10 successive inference requests reflected an edit to the prompt. To roughly understand the scale of these edits, we calculated sequence similarity between successive prompts using difflib \cite{difflib}, specifically the \textit{SequenceMatcher} ratio that measures the similarity between two sequences by comparing the length of their longest matching sub-sequence to the total length of both sequences. The distribution of prompt similarity ratios (see figure \ref{fig:edit_ratios}) indicates that the majority of changes were roughly limited to a band between 0.7 and 1.0 similarity (1.0 being an exact match), reflecting the iterative nature of prompt engineering. Users made limited sequential changes as they pursued their task. However, a minority of edits also clustered closer to 0 (no matching sub-strings) indicating that more or less, the entire prompt was changed. One factor to explain  large changes to successive prompts is that users could potentially send requests from multiple platform client instances, which is not reflected in the raw data. As such a user could work on more than one prompt in parallel, which looks in the data stream like large successive changes.

\begin{figure}[H]
    \centering
    \includegraphics[width=0.4\textwidth]{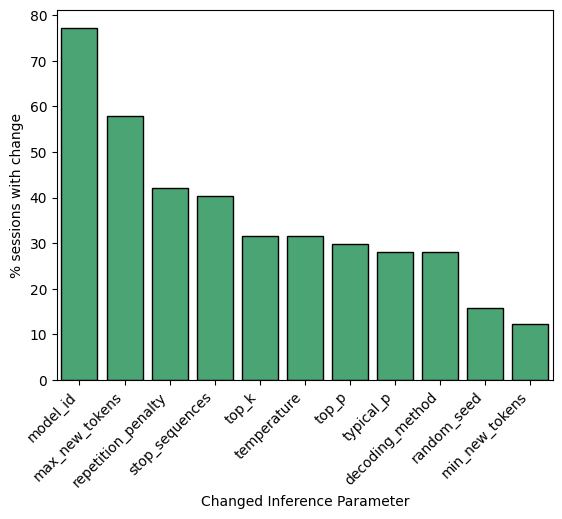}
    \caption{The occurrence of parameter changes as a percentage of sessions in which the change was observed. Users primarily changed the target language model, the maximum number of tokens to generate, and repetition penalty. Stop sequence, temperature and decoding method were also commonly changed.}
    \label{fig:parameter_changes}
\end{figure}

To understand what was happening when users did not make any edits to a prompt (and submitted an inference request), we also looked at model and parameter changes. Users regularly changed inference parameters when working on their prompts. 93\% of observed sessions involved one or more inference parameter change. The most commonly changed parameter was the target language model (\textit{model\_id}), followed by the \textit{max\_new\_tokens} and \textit{repetition\_penalty} parameters. See figure \ref{fig:parameter_changes} for a distribution of parameter changes observed in the data.



\begin{figure}[H]
    \centering
    \includegraphics[width=0.4\textwidth]{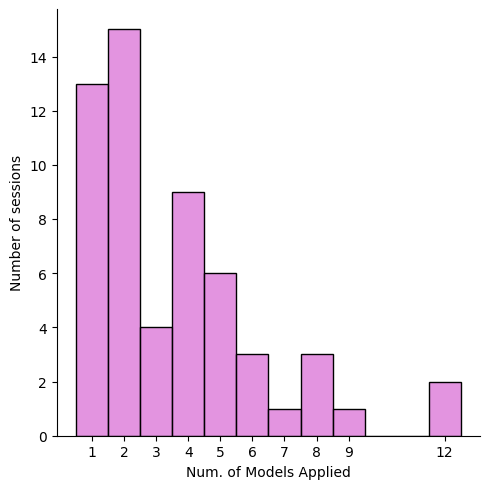}
    \caption{The number of models used per session.}
    \label{fig:model_counts}
\end{figure}

Given the frequency of change to the target model, we wanted to also understand how many different models were used within prompt editing sessions (See fig. \ref{fig:model_counts} for a distribution). The average number of models used in a typical session was 3.6 (SD=2.7), with a median of 3, indicating that users tend to work with multiple models, rather than focusing on a single one. Of course, this phenomenon is facilitated by the availability of a library of models within the observed prompting environment, and the ease of switching amongst these models when developing prompts. However the observed proclivity towards multiple models suggests that support for comparing model performance and better understanding model capabilities may help users to more efficiently work with LLMs.

\subsection{Prompt Editing Practices 
}
\subsubsection{Use cases}
The analyzed prompts reflected a wide variety of generative AI use cases. We found that use cases included: code/SQL generation (12), Q\&A/chat (11),  classification (7), extraction (7), summarization (5), reasoning (4), code explanation (3), JSON generation (2), and other (6). 

\subsubsection{Frequency of Prompt Component and Edit Types}
\begin{figure}[H]
    \centering
    \includegraphics[width=0.4\textwidth]{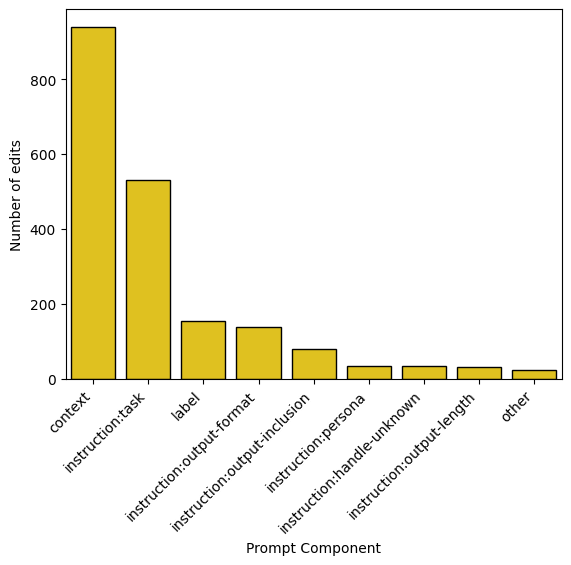}
    \caption{Number of edits that focused on each of the prompt components. Users primarily edited context, and task instructions to a lesser extent.}
    \label{fig:prompt_parts}
\end{figure}

Users focused primarily on editing task instructions and working with context (see Table~\ref{tab:code-part-freq} for descriptions and examples). 
Other notable foci were labels, output formatting, output inclusion, handling of unknowns and instructions related to controlling the generated output length (see Fig.~\ref{fig:prompt_parts}). The focus on editing of context over instruction was surprising, as well as the frequency of edits to the labels, compared to other types of instruction edits. 
 
\begin{figure}[tbh]
    \centering
    \includegraphics[width=0.4\textwidth]{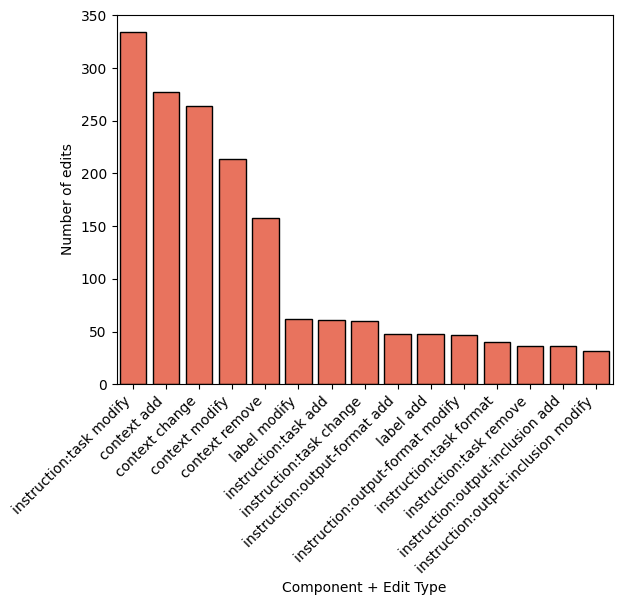}
    \caption{Number of edits of \textit{\{prompt component + edit type\}} pairs. For readability this figure is limited to the top 15 (of 41) most common pairs.}
    \label{fig:combined_edits}
\end{figure}

The most common type of edit (see Table~\ref{tab:code-change-freq} for definitions) applied was modification of the prompt in which the general meaning of the prompt remains consistent, followed by additions, changes where the meaning does not stay the same, removals and formatting (see Table~\ref{tab:code-change-freq} for descriptions and frequencies of edit types). 
Combining the prompt component with the type of edit applied gives a much clearer picture of the nature of the observed prompt editing activity (See fig. \ref{fig:combined_edits} for a distribution). The most common combination of  \textit{\{prompt component + edit type\}} pairs includes instruction tasks modification, context addition, change,  modification, and removal, and label modification. 



\subsubsection{Multiple Edits}
Edits were not always applied individually, we found that 22\% of all edits were \textit{multi-edits} meaning that the user made multiple simultaneous edits before resubmitting the prompt for inference. In those instances where multiple edits occurred simultaneously, the average number was 2.29 (SD=.59), with a median of 2. We suspected that these edits might include significant overlap with context, as users may have made edits to the instructions or labels, while also switching the document or query. We did find that 68\% of multi-edits included at least one context edit.  Nearly half (45\%) of multi-edits included both a context edit and an instruction edit, while 11\% of multi-edits included a context edit along with a label edit. While multi-edits may seem to increase efficiency, by attempting to fix multiple issues at once or by testing a new piece of context with new instructions or labels, making multiple edits at once could make tracking the impact an edit had on the generated output more challenging. Moreover we found that 1/5 edits were also accompanied by an inference parameter change. This type of behavior indicates the need for more systematic approaches to making changes and track the corresponding changes in model behavior, helping users to arrive at conclusions in a more comprehensible manner.  


\subsubsection{Rollbacks}
11\% of prompt edits undid or redid a previous edit (but are still counted as individual edits). Undoing or redoing a prompt edit could indicate a challenge in humans' ability to remember outcomes of previous attempts, or lack of certainty in the kinds of edits that might improve the output.
Interestingly, we found that prompt components that were edited less frequently often had higher percentages of un-done edits. 
We found high proportions of rollbacks for some prompt component edits, like 40\% of edits for instructions:handle-unknown were rolled back, as well as 25\% of instruction:output-length, 24\% of label edits, and 18\% of instruction:persona edits, compared to only 8-9\% of edits to context, instruction:task, and instruction:output-format. This may begin to explain why some of these components were edited less frequently- users may have found that their edits to those components did not have big impacts or had adverse or unexpected impacts. 

\subsubsection{Context}
We observed that the majority of the analyzed prompts/use-cases were context-based, meaning that input, grounding data, or examples were embedded within the prompt, distinct from the task instructions. In fact, context was the most edited component across all of the analyzed sessions (see figure \ref{fig:prompt_parts}) reflecting the importance of this construct for enterprise tasks. 

We observed two common patterns of context additions: 1) simulating dialog, and 2) adding examples. The prompt development interface our users worked with appended generated output directly to the input prompt text. This design made it very easy to generate turns in a conversation or dialog. Because each of these turns is a new generation, this accumulates a significant number of context additions. Further, much of this context would need to be removed for each of the other kinds of edits a user might make, before testing the conversation again. This kind of behavior matches Q\&A use cases the best, but can also be applicable in other generation tasks where a user might want to experiment with the length of the output to see what else a model would generate. Approximately 19\% of our prompts were Q\&A use cases.
Another kind of context addition was examples, which users added and removed, sometimes individually. 
Because examples can have significant effects on the generation, users may have wanted to add in examples individually or add and remove them to compare their impacts. 

Replacing and modifying existing context were also very common activities. We noticed that users would use a particular context to develop and refine their task instructions, and then proceed to evaluate the robustness of the instructions by switching in and out alternate contexts and observing model output. It was also common for users to directly edit existing context such a making changes to contextual queries or input data examples. 



\subsubsection{Instructions}
The editing behaviors of the task instructions we observed supports the common intuition that prompt development is an iterative trial and error process. 
The most common type of instruction edit was a task modification. In these situations, users were often re-wording the description of what the LLM should generate in order to improve the output. Instruction modifications could be as small as changing the capitalization, changing between a written word for a number and the numerical representation, or changing the punctuation. Some more complex ways users rephrased task instructions included: rephrasing an instruction between a statement and a question, rephrasing an instruction from a command to a description of what will be generated next, adding more detail, and simplifying. After modifying task instructions, the next most common activity was adding new task instructions, which often included specific rules or methods to generate the output. Somewhat surprisingly, edits to other types of instructions (output-format, output-inclusion, persona, handle-unknown, and output-length) were uncommon. Adding or modifying instructions to define the output format and what should be included were of the most common of these edits. One reason for this may be the types of use cases we saw in these editing sessions. For Q\&A and code generation use cases, the output format and inclusion criteria may be more standardized and straightforward than in less common use cases like JSON generation, classification, and extraction.


\subsubsection{Labels}
Edits in the labels of the prompt were the third most common prompt component edits. Label edits typically took the form of an identifier followed by a colon, a separator, or a start/end tag which delineates a section of the information. Prompts used labels throughout, such as for the instructions, context, examples, dialog, and output. Label modifications are likely driven by users' desire to have the LLM focus on certain potentially parameterize-able constructs within the prompt. Editing the labels for the output was also a common occurrence, as the label for output is the text directly prior to the generated text. In some cases, the output label acted like an instruction or a repetition of the instructions, indicating what should directly follow it. Since the output label has a potentially important role in the generation, users may have edited it to try to gain more control over the output. 







\section{Limitations}
Our data was anonymous, thus limiting our knowledge about the users' previous experience with prompting or their role in the organization. Thus, we cannot draw conclusions about how prompting practices may differ between novices and experts, nor along any other demographic dimensions. We sampled broadly to attempt to mitigate this limitation, as our user population has a wide range of skills and backgrounds. Further, due to the uneven distribution of use cases in prompt development, our sample does not cover use cases evenly. This may impact the frequency of edit types. We are also unable to share exact examples of prompt changes due to privacy restrictions on the data, limiting our ability to show what these prompts and edits look like in practice.

\section{Design Implications and Future Work}




\subsubsection{Prompt Debugging and Testing Support}
Our work explores prompt editing practices in a prompting tool, where users iterate on their prompt until they feel that it is good enough to move it to code or more in-depth evaluation. However, behaviors like undoing and redoing the same changes, making multiple edits at a time, frequent context changes, and model and parameter edits indicate inefficiencies in the iterative process of prompt engineering. 

Undoing and redoing the same edits could indicate challenges users have remembering how the output changed for different prompt edits, or that they have already made that change. This indicates a need for further support to reduce the cognitive load on users, such as by capturing  edits and corresponding outputs in more systematic and accessible ways, so users don't need to remember or re-run the same prompt multiple times. However, a version history that keeps track of every edit could also be hard to keep track of. Researchers~\cite{bach2022promptsource,mishra2023promptaid,strobelt2022interactive} and commercial systems (one of which is promptfoo~\footnote{https://www.promptfoo.dev/}, but many more exist and are not all publicly available) have been developed to support prompt iteration and comparison. These systems are often graphical user interfaces, rather than code that could use version control such as Git, but often lack sophisticated version control features that address the particular challenges we found in prompt iteration.

Making multiple edits at once may make it more difficult to determine the impacts of a particular edit, leading to further edits to isolate the effective or ineffective edit. We often found that in cases of multiple edits, context edits happened alongside other kinds of edits.
Some prompt engineering tools support multiple values for variables within a prompt, enabling multiple versions of contexts~\cite{arawjo2023chainforge,strobelt2022interactive}. We also found that users were often adding and removing parts of the context, which may lend itself to a different type of support, such as the ability to easily comment out or disable parts of a prompt without deleting them.


\subsubsection{Prompt Structure}
Both edits to the prompt label and formatting edits, which include white space and punctuation changes, point to attempts to structure prompts. Labels and complex structure may be more important and widely used in business contexts, where prompts are often grounded with documents, include examples, and may involve user and agent conversation. Further, while components like prompt instructions and context may be highly use case specific, structuring of prompts might be more easy to standardize. Some potential ways to support prompt engineering may be programming frameworks like LangChain\footnote{https://python.langchain.com/} and Microsoft Guidance\footnote{https://github.com/guidance-ai/guidance} or visual environments~\cite{wu2022promptchainer}. However, these tools focus more on the construction and connection of multiple prompts or prompt components, rather than supporting a user in deciding on a particular prompt structure. Other research has considered specific elements that can be added to augment a prompt, like instance-specific ``hints''~\cite{li2024guiding}. We suggest that future research investigate the potential for more structure to support users from both the model and the user interface perspectives and explore whether or not this structure can reduce prompt engineering effort.

\section{Conclusion}
In this work, we present an analysis of 57 prompt editing sessions, including 1523 individual prompts, from an enterprise LLM tool that supports prompt experimentation and development with LLMs. We found that users are often editing their prompt in addition to, or instead of, editing other model parameters. Many of these edits are quite small, likely indicating a tweak or iteration to one prompt rather than swapping in an entirely different prompt or task. Our qualitative analysis of edits shows that users are most often modifying the prompt context, including examples, grounding documents, and input queries. We were surprised that context edits outnumbered instruction edits, in which users are describing the task that they need accomplished, or particular elements of the task like the format of the output, what should or shouldn't be in the output, the output length, or the persona. Further, edits to labels, which define the components of the prompt, were also common. These findings provide critical insights into existing prompt editing practices and inform future directions on how to support prompt engineering in more effective ways. We are currently exploring various forms of prompt engineering assistance informed by the results of this study, such as for example, structured prompting, prompt histories \& comparisons, variation authoring, semi-automated explorations of variations, and built-in prompt quality evaluation based on use case-specific metrics.

\appendix



\bibliographystyle{named}
\bibliography{ijcai24}

\end{document}